\documentclass[%
reprint,
amsmath,amssymb,
aps,
]{revtex4-1}

\usepackage{graphicx}
\usepackage{dcolumn}
\usepackage{bm}
\usepackage[colorlinks=true,linkcolor=red,citecolor=blue]{hyperref}
\usepackage[mathlines]{lineno}

\usepackage{bbm}
\usepackage{amsmath}
\usepackage{verbatim}
\usepackage{graphicx}
\usepackage{braket,amsmath}
\usepackage[dvipsnames, svgnames, x11names]{xcolor}
\usepackage{array}
\usepackage{bibunits}
\usepackage{braket}
\usepackage{mathrsfs}
\usepackage{amsmath}
\usepackage{graphicx}
\usepackage{dcolumn}
\usepackage{bm}
\usepackage[capitalise]{cleveref}
\crefname{section}{Sec.}{Secs.}
\crefname{appendix}{App.}{Apps.}

\begin{document}
	
	\preprint{APS/123-QED}
	
	\title{Unitary and efficient spin squeezing in cavity optomechanics}
	
	\author{Lei Xie}%
	\affiliation{%
		Department of Physics, Wenzhou University, Zhejiang 325035, China
	}%

	\author{Zhiqi Yan}%
	\affiliation{%
		Department of Physics, Wenzhou University, Zhejiang 325035, China
	}%

	\author{Lingxia Wang}%
	\affiliation{%
	Department of Physics, Wenzhou University, Zhejiang 325035, China
	}%

	\author{Di Wang}%
	\affiliation{%
		Department of Physics, Wenzhou University, Zhejiang 325035, China
	}%

	\author{Jinfeng Liu}%
	\affiliation{%
		Department of Physics, Wenzhou University, Zhejiang 325035, China
	}%
	
	\author{Yiling Song}
	\affiliation{%
		Department of Physics, Wenzhou University, Zhejiang 325035, China
	}%
	
	\author{Wei Xiong}
	\email{xiongweiphys@wzu.edu.cn}
	\affiliation{%
		Department of Physics, Wenzhou University, Zhejiang 325035, China
	}%
	\author{Mingfeng Wang}
	\email{mfwang@wzu.edu.cn}
	\affiliation{%
		Department of Physics, Wenzhou University, Zhejiang 325035, China
	}%
	
	
	\date{\today}
	
	\begin{abstract}
		We propose an approach to produce spin squeezed states of a large number of nitrogen-vacancy centers in diamond
		nanostructures coupled to an optical cavity. Unlike the previous squeezing method proposed by Bennett \emph{et al.} [\href{https://doi.org/10.1103/PhysRevLett.110.156402}{Phys. Rev. Lett. 110, 156402 (2013)}], which is limited by phonon number fluctuations due to the existence of phonon-spin entanglement, our proposal can completely erase the entanglement between  spins and hybrid phonon-photon mode mediating the effective spin-spin interaction, and thus achieves unitary one-axis-twisting interactions between nitrogen-vacancy centres, yielding a squeezing scaling $J^{-2/3}$, where $J$ is the total angular momentum. We found that, under certain conditions, our method has the potential to enhance the spin-spin nonlinear interactions. We also proposed a scheme utilizing repeatedly applying the one-axis-twisting evolution to two orthogonal spin directions, which enables the transformation of the one-axis-twisting interactions into two-axis-twisting type, and therefore leads to the spin squeezing with Heisenberg-limited scaling $J^{-1}$. Taking into account the noise effects of spin dephasing and relaxtion, we found that the proposed approaches are robust against imperfections.
		\begin{description}
			\item[PACS numbers]
			 71.55.-i, 07.10.Cm, 42.50.Dv
		\end{description}
	\end{abstract}
	
	
	\pacs{Valid PACS appear here}
	\keywords{Suggested keywords}
	\maketitle
	
	
	\section{\label{sec:level1}INTRODUCTION}
	Spin squeezed states (SSSs) \cite{PhysRevA.47.5138,PhysRevA.50.67,PhysRevA.49.4968,PhysRevA.81.021804,PhysRevLett.110.120402,PhysRevA.96.050301,PhysRevLett.121.070403} are currently attracting particular interests because they are key ingredient for high-precision measurements \cite{PhysRevLett.82.4619,PhysRevLett.86.5870,PRXQuantum.3.010354,PhysRevX.10.031003,PhysRevLett.125.200505} and quantum information science \cite{PhysRevLett.86.4431,PhysRevA.68.012101,PhysRevLett.95.120502,PhysRevLett.102.100401,PhysRevLett.119.193601,PhysRevLett.125.143601,PhysRevLett.126.060401}.
	Quantum correlations of SSSs enables achieving the sensitivity of precision measurements beyond the standard quantum limit (SQL) \cite{RevModPhys.90.035005}. In particular, SSSs have been used to increase the sensitivity of atomic clocks \cite{PhysRevA.50.67}, magnetometry \cite{PhysRevLett.104.093602,PhysRevA.85.022125}, and improve the fidelity of quantum memory \cite{PhysRevA.74.011802}.
	
	By now, various spin-squeezing approaches have been proposed and experimentally demonstrated, including the transformation of squeezed light onto atoms \cite{PhysRevLett.83.1319}, projection through quantum nondemolition (QND) measurement \cite{appel2009mesoscopic,PhysRevLett.104.073604,PhysRevLett.106.133601}, steady-state squeezing by reservoir engineering \cite{PhysRevLett.110.120402,PhysRevLett.127.083602}, and unitarily evolving spin system via spin-spin nonlinear interactions \cite{PhysRevA.47.5138}.  This last method is probably the most widely studied one to date. Depending on the form of the interaction Hamiltonian, two major nonlinear interactions, one-axis-twisting (OAT) \cite{PhysRevLett.104.073602,riedel2010atom,gross2010nonlinear,hosten2016quantum} and two-axis-twisting (TAT) \cite{PhysRevA.80.032311,PhysRevLett.107.013601,borregaard2017one,PhysRevLett.125.203601,PhysRevA.96.013823,bao2020spin} interactions, have been proposed to produce SSSs. Besides the spin squeezing, the OAT interactions have also found important applications in detection-noise evading \cite{PhysRevLett.116.053601}, Greenberger-Horne-Zeilinger (GHZ) states preparation \cite{PhysRevLett.82.1835}, and continuous-variable quantum computation performance \cite{PhysRevLett.119.010502}. Therefore, in recent years a great effort has been oriented to the realization of OAT interaction in different physical platforms, such as trapped ions \cite{PhysRevLett.82.1835}, Bose-Einstein condensates \cite{gross2010nonlinear}, room-temperature (or hot) atoms \cite{PhysRevA.96.013823}, and nitrogen-vacancy (NV) centers in diamond \cite{PhysRevLett.110.156402}.
	
	NV centers in diamond are one of the most promising solid-state systems and have motivated extensive research due their long coherence times and capability for optical addressability \cite{jelezko2004observation,balasubramanian2009ultralong,santori2006coherent,buckley2010spin,dutt2007quantum,doherty2013nitrogen,PhysRevB.79.041302}. The ability to control interactions between various NV centers is highly desirable in the field of quantum-gates operation and quantum-enhanced sensing. Possible routes towards this goal include the NV centers mediated by photons \cite{faraon2011resonant,englund2010deterministic,PhysRevLett.110.243602} or phonons \cite{kolkowitz2012coherent,arcizet2011single,rabl2010quantum,PhysRevLett.116.143602}. Among the later approaches, an elegant one, proposed by Bennett\emph{ et al.} \cite{PhysRevLett.110.156402}, is based on coupling between NV centers and a vibrational mode of a diamond resonator. Such a coupling is achieved by embedding NV centers into a diamond nanobeam, and the vibration of nanobeam yields a strain-induced vibrating electric field \cite{PhysRevB.85.205203}, which causes the transition of the electronic ground state of the NV centers and therefore effectively realizes the coupling of a single resonant mode of the nanobeam to the NV centers. Using the phonon of the resonant mechanical mode as an intermediary, it is possible to achieve an effective OAT interaction and thus spin squeezing of an NV ensemble \cite{PhysRevLett.110.156402}. This approach, however, has the limitation that the spin system after the OAT interaction is still entangled with the phonons. If such spin-phonon entanglement is ignored completely, it will cause \emph{non-unitary} evolution in the spin subspace and add an undesirable dephasing to the spin system \cite{PhysRevA.85.013803}, reducing the achievable spin squeezing. Although one can disentangle the spins and the phonons via decoupling operations \cite{PhysRevLett.88.207902,PhysRevA.90.042118,PhysRevA.58.2733,PhysRevA.92.013825,viola1999dynamical}, which on the one hand complicates the implementation of spin squeezing and on the other hand may make the scheme unsuitable for other quantum optics tasks, such as the GHZ state creation \cite{PhysRevLett.82.1835}.
	
	In this paper, we propose an alternative approach to realize \emph{unitary} OAT evolution of an NV ensemble embedded in a nanobeam. Our approach is based on the coupling between NV centers and a cavity-optomechanics system. We show that, by simply coupling the nanobeam to an additional cavity mode, one is able to transform the Tavis-Cummings type interaction \cite{PhysRev.170.379} between NV spins and a mechanical mode as in Ref. \cite{PhysRevLett.110.156402}  into the Dicke type interaction \cite{PhysRev.93.99} between NV spins and a hybrid optical-mechanical mode. This Dicke type interaction allows for unitary OAT interaction induced by virtual photon-phonon processes, without leaving a trace of the spin state in the optical and mechanical modes. Consequently, the performance of the mechanical-strain-induced spin squeezing can be improved beyond the limit set by the phonon-number fluctuation. We also find that under certain circumstances our proposed scheme also exhibits the potential to realize much more stronger spin-spin nonlinear interactions. Moreover, by adding a time-dependent rotation to the collective spin during the Dicke-type interaction, we also convert the effective OAT interaction into the much more efficient TAT interaction, which enables the achievement of spin squeezing near the Heisenberg limit \cite{RevModPhys.90.035005}. Taking into account the noise effects, we show that the proposed schemes are fairly robust against decoherences.
	
	The paper is organized as follows. In \cref{sec:level2} we introduce the spin system and give the interaction Hamiltonian of the hybrid system. In \cref{sec:level2.2} we analytically derive the effective OAT interaction. In \cref{sec:level2.3} a TAT squeezing scheme is presented. In \cref{sec:level3} the influence of noises is considered, and we also calculate the amount of spin squeezing in the presence of decoherences. Finally, we summarize in \cref{sec:level4}.
	\begin{figure}[t]
		\centering
		\includegraphics[scale = 0.45]{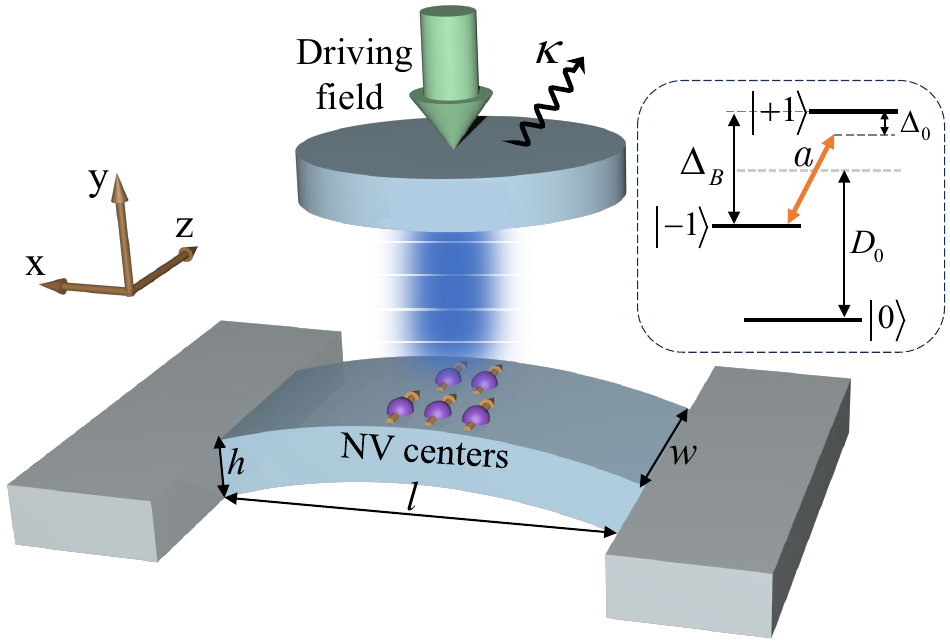}
		\caption{Schematics of the hybrid spin-cavity-optomechanical system.
			A collection of NV centers (purple arrows) weakly couples to a mechanical resonator via crystal strain, and at the same time the mechanical resonator also couples to a driven cavity through radiation pressure.  Inset: the energy level structure of the NV electron ground state. With a magnetic field $B$ applied along $ z $ direction, it causes a Zeeman splitting of the sublevels $ \left|{m_s} = {\pm 1} \right\rangle $, which couples off-resonantly to the mechanical mode $a$ with a detuning $\Delta_0$.}
		\label{fig1}
	\end{figure}
	
	\section{\label{sec:level2}The model}
	We consider a hybrid quantum system consisting of an NV ensemble weakly coupled to a mechanical resonator driven by an optical cavity. The negatively charged NV centers are near-surface embedded in a mechanical resonator with dimensions ($l$, $w$, $h$), as shown in \cref{fig1}. The electronic ground state of the NV center is a triplet with spin $S=1$, i.e., $\left|{m_s} = {0, \pm 1} \right\rangle$, as depicted in the inset of \cref{fig1}. The zero-field split between energy levels $\left| 0 \right\rangle$ and $\left| { \pm 1} \right\rangle$ is equal to ${D_0} = 2.87$ GHz \cite{PhysRevLett.113.020503,PhysRevLett.105.210501}. If an additional magnetic field is applied to the NV center, the degeneracy between energy levels $\left| { \pm 1} \right\rangle$ is lifted, resulting in a Zeeman splitting ${\Delta _B} = {g_e}{\mu _B}{B_z}$ ($\hbar=1$), where ${g_e}=2$ represents the Land\'{e} $g$-factor, ${\mu_B} = 13.996$ GHz/T denotes the Bohr magneton, and $B_z$ refers to the strength of the additional applied magnetic field. The interaction between each NV center and the diamond mechanical resonator can be described by ${H_i} = {g}\left( {\sigma _i^ + a + {a^\dag }\sigma _i^ - } \right)$ \cite{PhysRevLett.110.156402}, where $\sigma _i^ \pm  = {\left| { \pm 1} \right\rangle _i}\left\langle { \mp 1} \right|$ are the Pauli operators for the $i$th NV center, $a$ is the annihilation operator for the resonant mechanical mode at frequency ${\omega _a}$, and $g$ is the coupling strength. For a collective NV ensemble, it is necessary to introduce the collective spin operators, namely, ${J_z} = \sum\nolimits_{i=1}^N ({{{\left| 1 \right\rangle }_i}\left\langle 1 \right|}  - {\left| { - 1} \right\rangle _i}\left\langle { - 1} \right|)/2$ with $N$ being the number of spins and ${J_ \pm } = {J_x} \pm i{J_y} = \sum\nolimits_i {\sigma _i^ \pm }$, which satisfy the general angular momentum relations $[J_x,J_y]=iJ_z$. As a result, the total Hamiltonian of the setup presented in \cref{fig1} reads 
	\begin{eqnarray}
		H = {H_0} + {H_I} + {H_D},\label{eq1}
	\end{eqnarray}
	with
	\begin{subequations}
		\label{eq:whole}
		\begin{eqnarray}
			{H_0} &=& {\omega _a}{a^\dag }a + {\omega _b}{b^\dag }b + {\Delta _B}{J_z},\label{eq2a}\\
			{H_I} &=& {g}\left( {{a^\dag }{J_ - } + a{J_ + }} \right) - {g_0}{b^\dag }b\left( {{a^\dag } + a} \right),\label{eq3b}\\
			{H_D} &=& {\Omega _d}{b^\dag }\exp \left( { - i{\omega _d}t} \right) + \Omega _d^ * b\exp \left( {i{\omega _d}t} \right).\label{eq4c}
		\end{eqnarray}
	\end{subequations}
	Here, $H_0$ is the free Hamiltonian of the uncoupled system with ${\omega _b}$ being the frequency of the cavity mode $b$ and $\Delta_B$ being the contribution of the magnetic field $B_z$. $H_I$ is the Hamiltonian of the interaction between the systems, and $g_0$ is the coupling strength of the interaction between the cavity mode and the mechanical mode \cite{law1995interaction, PhysRevLett.114.093602}. ${H_D}$ denotes the Hamiltonian of driving field, with a frequency ${\omega _d}$ and an amplitude ${\Omega _d}$,  acting on the cavity mode. Below, we will show how to obtain the spin-spin nonlinear interactions based on the full Hamiltonian (\ref{eq1}).
	
	\section{\label{sec:level2.2} One-axis twisting spin-spin nonlinear interaction}
	Assume that the cavity-mechanical system is driving by a strong coherent field, and thus one can linearize the radiation-pressure Hamiltonian $\propto {b^\dag }b({a^\dag } +a )$. More specifically, the operators $a$ and $b$ can be expressed as the sum of their mean values and the corresponding fluctuations, such that $a=\langle a\rangle+\delta a$ and $b=\langle b\rangle+\delta b$, which enables us to determine the steady-state mean values of the mechanical mode $ a $ and the cavity mode $ b $ via the input-output theory \cite{walls1994gj}, resulting in $\langle a \rangle  =  - {g_0}{ | {\langle b \rangle } |^2}/(i{\gamma _a} - {\omega _a})$ and $\langle b \rangle  =  - {\Omega _d}/({\Delta _{bd}} - i\kappa) $, where ${\Delta _{bd}} = ( {{\omega _b} - {\omega _d}} ) - {g_0}( {\langle a \rangle  + {{\langle a \rangle }^ * }} )$ and $\gamma_a$ ($\kappa$) denotes the decay rate of the mechanical (cavity) mode. As a result, we obtain the linearized Hamiltonian for the hybrid quantum system \cite{PhysRevB.103.174106}
	\begin{eqnarray}
		{H_{{\rm{lin}}}} &=& {\Delta _B}{J_z} + {\omega _a}\delta {a^\dag }\delta a + {g}\left( {\delta {a^\dag }{J_ - } + \delta a{J_ + }} \right)  \nonumber\\
		&& + {\Delta _{bd}}\delta {b^\dag }\delta b - G\left( {\delta b + \delta {b^\dag }} \right)\left( {\delta a + \delta {a^\dag }} \right),\label{eq3}
	\end{eqnarray}
	where the last term represents the linearized radiation-pressure Hamiltonian with $ G = g_0 \sqrt{|\langle b\rangle|^2}  $ being the linearized cavity-mechanical coupling strength. Now, $ G $ contains the effect of the driving field, and therefore the coupling between the cavity and mechanical modes is greatly enhanced, while the quantum behavior of the system is characterized by the fluctuation operators $ \delta a $ and $ \delta b $ in Eq. (\ref{eq3}).
	\begin{figure}[t]
		\centering
		\includegraphics[scale=0.5]{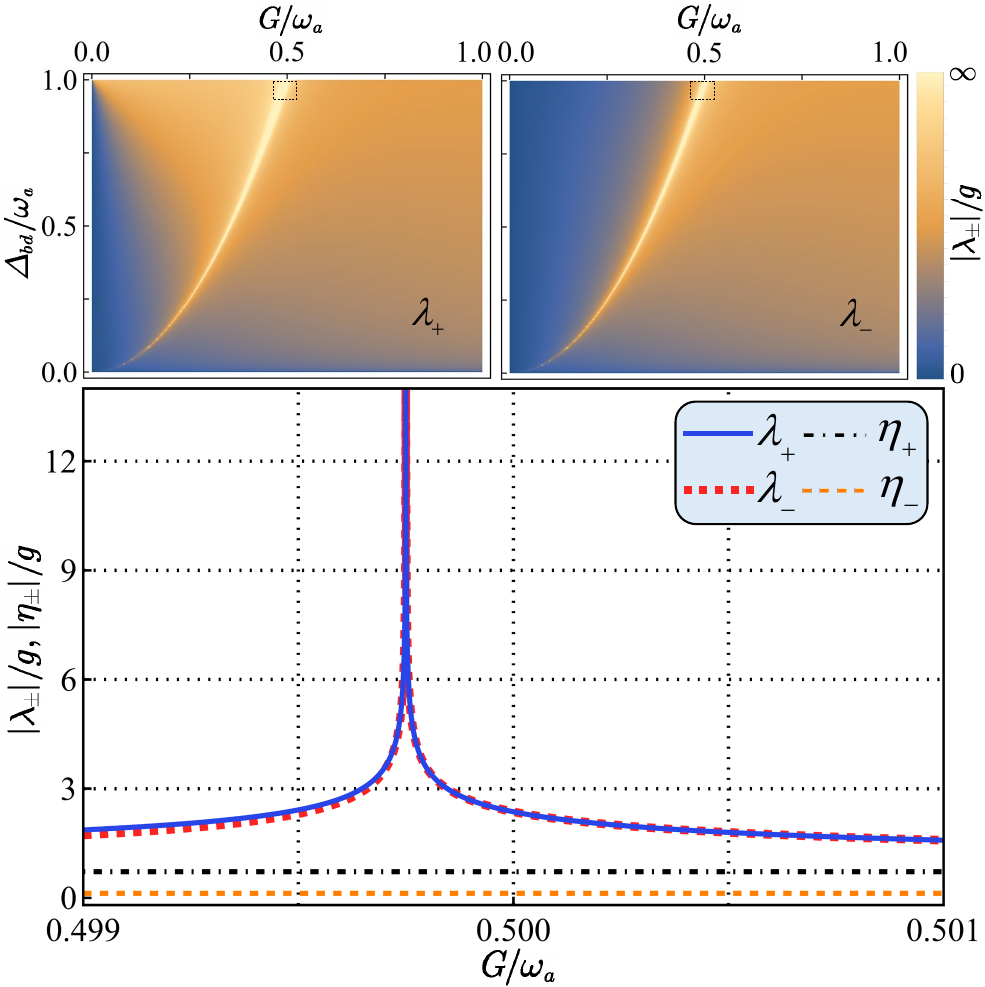}
		\caption{Density plot of $\lambda_+$ (upper left) and $\lambda_-$ (upper right) as a function of the effective detuning $\Delta_{bd}$ and the coupling strength $G$.  The boxes (dashed rectangle) highlight the parameter regimes ($\Delta_{bd}\to{\omega _ a }$, $G\to{\omega _ a }/2$) that are relevant for our protocol. The bottom plot shows
the coupling strength  $\lambda_\pm$($\eta_\pm$) between the NV centers and the Bogoliubov mode $B_-$ ($B_+$) versus $G$ for the case $(\omega_a-\Delta_{bd})/\omega_a=10^{-3}$.}
		\label{fig2}
	\end{figure}

	The cavity-mechanical Hamiltonian of Eq. (\ref{eq3}) can be diagonalized in terms of the Bogoliubov modes \cite{lu2013quantum,PhysRevA.86.012316}
	\begin{eqnarray}
		{B_ - } &=& \frac{1}{2}\frac{{\cos \theta }}{{\sqrt {{\Delta _{bd}}{\omega _ - }} }}\left[ {\left( {{\omega _ - } + {\Delta _{bd}}} \right)\delta b + \left( {{\omega _ - } - {\Delta _{bd}}} \right)\delta {b^\dag }} \right]\nonumber\\
		&&- \frac{1}{2}\frac{{\sin \theta }}{{\sqrt {{\omega _a}{\omega _ - }} }}\left[ {\left( {{\omega _ - } + {\omega _a}} \right)\delta a + \left( {{\omega _ - } - {\omega _a}} \right)\delta {a^\dag }} \right],\nonumber\\
		{\rm{     }}{B_ + } &=& \frac{1}{2}\frac{{\sin \theta }}{{\sqrt {{\Delta _{bd}}{\omega _ + }} }}\left[ {\left( {{\omega _ + } + {\Delta _{bd}}} \right)\delta b + \left( {{\omega _ + } - {\Delta _{bd}}} \right)\delta {b^\dag }} \right]\nonumber\\
		&&+ \frac{1}{2}\frac{{\cos \theta }}{{\sqrt {{\omega _a}{\omega _ + }} }}\left[ {\left( {{\omega _ + } + {\omega _a}} \right)\delta a + \left( {{\omega _ + } - {\omega _a}} \right)\delta {a^\dag }} \right],\nonumber
	\end{eqnarray}
	where $ \tan 2\theta  = {{4G\sqrt {{\Delta _{bd}}{\omega _a}} } \mathord{\left/{\vphantom {{4G\sqrt {{\Delta _{bd}}{\omega _a}} } \left( {\Delta _{bd}^2 - \omega _a^2}\right) }} \right.\kern-\nulldelimiterspace} \left( {\Delta _{bd}^2 - \omega _a^2}\right) }$. The normal modes $B_\pm$ satisfy $[B_-,B_-^\dag]=[B_+,B_+^\dag]=1,[B_-,B_+]=[B_-,B_+^\dag]=0$ and have the frequencies
	\begin{eqnarray}
		\omega _ \pm ^2 &=& \frac{1}{2}\left[ {\Delta _{bd}^2 + \omega _a^2} \right.\nonumber\\
		&&\left. { \pm \sqrt {{{\left( {\Delta _{bd}^2 + \omega _a^2} \right)}^2} + 4{\Delta _{bd}}{\omega _a}\left( {4{G^2} - {\Delta _{bd}}{\omega _a}} \right)} } \right],\label{eq4}
	\end{eqnarray}
	respectively. Defining the new coupling strength ${\lambda _ \pm } \equiv  - {g}\sin \theta ({\omega _a} \pm {\omega _ - })/2\sqrt {{\omega _a}{\omega _ - }} $ and ${\eta _ \pm } \equiv {g}\cos \theta ({\omega _a} \pm {\omega _ + })/2\sqrt {{\omega _a}{\omega _ + }}$, the diagonalization process yields
	\begin{eqnarray}
		\nonumber H_d &=& {\Delta _B}{J_z} + {\omega _ - }B_ - ^\dag {B_ - } + {\omega _ + }B_ + ^\dag {B_ + } \\
		\nonumber &+& {\lambda _ + }\left( {B_ - ^\dag {J_ - } + {B_ - }{J_ + }} \right) + {\lambda _ - }\left( {B_ - ^\dag {J_ + } + {B_ - }{J_ - }} \right) \\
		&+& {\eta _ + }\left( {B_ + ^\dag {J_ - } + {B_ + }{J_ + }} \right) + {\eta _ - }\left( {B_ + ^\dag {J_ + } + {B_ + }{J_ - }} \right),\label{eq5}	
	\end{eqnarray}
	which indicates that the collective spin couples simultaneously to the normal modes  $B_\pm$ via the isotropic Dicke interactions \cite{PhysRevA.75.013804,PhysRevLett.118.080601}  [since here $\lambda_+\neq\lambda_-$ and $\eta_+\neq\eta_-$ can be realized, it is called Dicke interaction when $\lambda_+=\lambda_-$ (or $\eta_+=\eta_-$)]. One can tailor the desired interactions of atoms with the normal-modes by appropriately tuning the values of $\Delta_{bd}$ and $G$. Choosing $\Delta_{bd}\to{\omega _ a }$ and $G\to{\omega _ a }/2$, we will have $\omega_+\simeq\sqrt{2}\omega_a\gg\omega_-$, resulting in $\lambda = {\lambda _ + } \simeq {\lambda _ - } =  - \sqrt {2{\omega _a}} {g}/4\sqrt {{\omega _ - }} $ and ${\eta _ \pm } = (1 \pm \sqrt 2 ){g}/{2^{7/4}}$.
	Moreover, if $\sqrt{\omega_a/\omega_-}\gg 1$, then $\lambda \gg {\eta _\pm }$, which enables us to safely neglect the $\omega_+$ mode, as shown in \cref{fig2}. As a result, Eq. (\ref{eq5}) can be approximated as
	\begin{eqnarray}
		H'_d \simeq {\omega _ - }B_ - ^\dag {B_ - } + {\Delta _B}{J_z} + 2\lambda \left( {B_ - ^\dag   + {B_ - }} \right)J_x,\label{eq6}
	\end{eqnarray}
	which is exactly the Dicke model \cite{PhysRev.93.99}. Next, we assume that the two energy levels $\ket{\pm 1}$ are degeneracy with $\Delta_B=0$ and ${\omega _ - } \gg 2\lambda$, the Hamiltonian $H'_d$ can be approximately diagonalized by the unitary transformation ${e^S}{H'_d}{e^{ - S}} = {H_{\rm{eff}}}$ with $S = {2\lambda}({B_ - ^\dag   - {B_ - }})J_x/{\omega_- } $. Truncating to the second order of $\lambda/\omega_-$, the effective Hamiltonian $H_{\rm{eff}}$ is reduced to the form
	\begin{eqnarray}
		H_{\rm{eff}} = {\omega _ - }B_ - ^\dag {B_ - } - \chi J_x^2.\label{eq7}
	\end{eqnarray}
	The second term of Eq. (\ref{eq7}) is exactly the OAT Hamiltonian \cite{PhysRevA.47.5138} with a Bosons-modes-mediated spin-spin
	coupling strength
	\begin{eqnarray}
		\chi  = \frac{{4{\lambda ^2}}}{{{\omega _ - }}} = \mathcal{A} \cdot \frac{{{g}}}{r},~~~~~~\mathcal{A} = {\left( {\frac{{{\omega _a}}}{{{2rg}}}} \right)^{1/3}},\label{eq8}
	\end{eqnarray}
	where we have defined $r=\omega_-/2\lambda$, which leads to $\omega_-=\sqrt[3]{{r^2}g^2{\omega _a}/2}$. Compared to the existing proposal \cite{PhysRevLett.110.156402}, in which the effective Hamiltonian is of the form: $ H_{\rm{eff}}\propto \delta a^\dag \delta aJ_z+\frac{g}{r} J_z^2$ \cite{footnote1}, our scheme offers many advantages. First, from Eq. (\ref{eq7}) we can see that the atomic state is completely decoupled from the photonic mode, as evidenced by the absence of the term $\delta a^\dag \delta aJ_z$. This implies that there is no need for decoupling operations, resulting in a simplified experimental implementation. Second, there exists an additional prefactor $\mathcal{A}$ in the coupling constant, which is proportional to the ratio between the frequency of the mechanical mode and the spin-mechanical-nanoresonator coupling strength. While, in reality, such a ratio is normally large, it will amplify the spin-spin coupling strength. As an example, choosing the parameters $\omega_a/2\pi \approx 1.88~\rm{GHz}$ and $g/2\pi\approx0.72~\rm{kHz}$ presented in Ref. \cite{PhysRevLett.110.156402}, the interaction strength is remarkably enhanced by a factor of $\mathcal{A}\approx24$ for $r=10^2$. The amplification factor can be further increased by enlarging the size of the nanooscillator. Since $\omega_a\propto h/l^2$ and $g\propto 1/(l^{3/2}w^{1/2})$ as was shown in Ref. \cite{PhysRevLett.110.156402}, we have $\omega_a/g\propto h(w/l)^{1/2}$. If the size of the nanooscillator is increased by a factor of ten [that is, $(l,w,h)\rightarrow (10l,10w,10h)$], one will have $\mathcal{A}\approx 50$. In short, our proposed scheme is particularly well-suited for quantum systems with large $\omega_a$ and small $g$ (which results in large $\mathcal{A}$).
	
	\section{\label{sec:level2.3} Two-axis twisting spin-spin nonlinear interaction}
	
	A more efficient approach to achieve spin squeezing would be the TAT interaction, which enables the reduction of Heisenberg-limited noise \cite{PhysRevA.47.5138}. However, implementing the TAT in realistic quantum systems is quite challenging \cite{PhysRevA.96.013823,bao2020spin}. Here, we present an alternative way to realize the TAT evolution in the NV ensemble. To do so, we introduce a time-dependent modulation to the energy separation between $\ket{+1}$ and $\ket{-1}$, which can be described by the time-dependent non-interacting Hamiltonian \cite{PhysRevA.96.043849}
	\begin{eqnarray}
		{H_M}\left( t \right) &=& \xi \nu \cos \left( {\nu t} \right){J_z},\label{eq9}
	\end{eqnarray}
	where $\xi$ denotes the modulation amplitude and $\nu$ represents the modulation frequency. Such modulation can be performed by applying a time-varying magnetic field to the NV centers along the $z$ axis. With this frequency modulation, the total Hamiltonian of Eq. (\ref{eq6}) becomes
	\begin{eqnarray}
		{H'_d}\left( t \right) &=& \left[ {{\Delta _B} + \xi \nu \cos \left( {\nu t} \right)} \right]{J_z} + {\omega _ - }B_ - ^\dag {B_ - } \nonumber\\
		&&+ 2\lambda \left( {B_ - ^\dag + {B_ - }} \right)J_x.\label{eq10}
	\end{eqnarray}
	Changing to a rotating frame with respect to $H_0(t)=\omega_{-} B_{-}^{\dagger} B_{-}+[\Delta_B+\xi \nu \cos (\nu t)] J_z$ by evaluating ${\tilde{H}'_d(t)} = V(t){H'_d}(t){V^\dag(t) }-iV(t)\dot{V}^\dag(t)$, where $V(t)= \mathcal{T}\exp [ { - i\int_0^t {{H_0}( \tau  )d\tau } }]$ with $\mathcal{T}$ being the time-ordering operator, we obtain
	\begin{eqnarray}
		\nonumber \tilde{H}'_d\left( t \right) &=&  \sum\limits_{n =  - \infty }^\infty  {{J_n}\left( \xi  \right)} {\lambda }\left( {B_ -  {J_ + }{e^{i\left( {\delta ^ -  + n\nu } \right)t}} + \rm{H.c.}} \right)\\
		\nonumber && + \sum\limits_{m =  - \infty }^\infty  {{J_m}\left( \xi  \right)} {\lambda  }\left( {B_ - ^\dag {J_ + }{e^{i\left( {\delta ^ +  + m\nu } \right)t}} + \rm{H.c.}} \right),\\\label{eq11}
	\end{eqnarray}
	where $ {\delta^\pm} = {\Delta _B} \pm {\omega _ - } $ are the detunings between the unmodulated two-level system and the normal mode. In the derivation of Eq. (\ref{eq11}), we have used the Jacobi-Angel expansion ${\exp[{i\xi \sin \left( {\nu t} \right)}]} =\sum\nolimits_{ n=- \infty }^{ + \infty } {{{J_n}\left( \xi  \right)} {\exp({in\nu t})}}$, where $ {J_n}\left( \xi  \right) $ is the $ n $th Bessel function of the first kind. Eq. (\ref{eq11}) indicates that the frequency modulations of the atomic system leads to the interactions of the atoms with a series of sidebands of the normal modes. The coupling between the sideband modes and the spins can be adjusted by tuning the parameters $\xi$ and $\nu$ \cite{PhysRevA.96.043849,PhysRevA.103.032601}.
	
	For the rotational interaction terms [the first line of Eq. (\ref{eq11})], the detunings of the $n$th sideband  $({n\nu  + {\delta ^-}}) $ and the $n'$th sideband $({n'\nu  + {\delta ^-}}) $ are separated from each other by $(n - n')\nu\geq\nu ~(n\neq n')$. When $\nu  \gg ({\lambda},\delta^-)$, the detuning of each non-zero-order sideband (${n\nu  + {\delta ^-}}$) is much larger than the zero-order sideband (${{\delta ^-}}$). If we focus solely on dynamics that are time-averaged over a period $\tau$ much longer than the period of the modulation ($\tau\gg 2\pi/\nu$), then all the (fast oscillating) interaction terms of the  non-zero-order sidebands can be neglected, keeping only the (slow oscillating) interaction term of the zero-order sidebands (such process is also known as the rotating-wave approximation). For the counter-rotation terms [the second line of Eq. (\ref{eq11})], there exists a counter-rotation sideband with the smallest detuning $\left| {{\Delta _{m_0}}} \right| = \min { {| {{\delta ^ + } + m\nu } |,m \in Z} }$ (with $Z$ being the integers). Under the conditions $\nu\gg |\Delta _{m_0}|$, only the counter-rotating term of the sideband $\Delta_{m_0}$ survives. As a result, the Hamiltonian of Eq. (\ref{eq11}) can be approximated as \cite{PhysRevA.96.043849}
	\begin{eqnarray}
		{\tilde H'_d}\left( t \right){\rm{ }} = {\lambda _0}{B_ - }{J_ + }{e^{i{\delta ^ - }t}} + {\lambda _{{m_0}}}B_ - ^{{{\dag}}}{J_ + }{e^{i{\Delta _{{m_0}}}t}} + {\rm{H}}.{\rm{c}},\label{eq12}
	\end{eqnarray}
	where $ {\lambda_0}={\lambda}{J_0}( \xi ) $ and $ {\lambda_{m_0}} = {\lambda} {J_{m_0}} ( \xi ) $, which can be tuned by varying the modulation amplitude $\xi$. The Hamiltonian (\ref{eq12}) describes an effective isotropic Dicke model (when $\lambda_0\neq \lambda_{m_0}$) with an effective frequency of the normal mode $\tilde\omega_-=(\Delta_{m_0}-\delta^-)/2$ and an effective frequency of the NV centers $\tilde\Delta_B=(\Delta_{m_0}+\delta^-)/2$. If we set ${\delta^-} = -\Delta_{m_0} \equiv  {\Delta}$, then $\tilde\Delta_B=0$, meaning that the quantum two-level system is effectively degenerate. Next, we apply the transformation ${V_1} = \exp [ i{{\Delta}B_ - ^\dag {B_ - }} t] $ to return the Hamiltonian $ {\tilde H'}_d (t) $ back to a non-rotating coordinate system and obtain
	\begin{eqnarray}
		{{\tilde H}_1} = \Delta B_ - ^\dag {B_ - } + \lambda \left( {B_ - ^\dag \Omega  + {B_ - }{\Omega ^\dag }} \right),\label{eq13}
	\end{eqnarray}
	where we have defined $\Omega  = {J_ - }{J_0}( \xi  ) + {J_ + }{J_{m0}}( \xi  ) $. Similarly, when the detunings ${\Delta} \gg \lambda > {\lambda_0},{\lambda_{m_0}}$, the Hamiltonian (\ref{eq13}) can be approximately diagonalized by using the transformation ${e^S}{{\tilde H}_1}{e^{-S}}$ with the generator $ S= {\lambda}(B_ - ^\dag {{\Omega} - {B_ - } \Omega^\dag }) /{\Delta}$, yielding
	\begin{eqnarray}
		\nonumber {H_{\rm{eff}}} &=& \Delta B_ - ^\dag {B_ - } - \frac{{\lambda _0^2 - \lambda _{{m_0}}^2}}{\Delta }\left( {2B_ - ^\dag {B_ - } + 1} \right){J_z} \\
		&& + \frac{{{{\left( {\lambda _0 - \lambda _{{m_0}}} \right)}^2}}}{\Delta }J_z^2 - \frac{{4\lambda _0\lambda _{{m_0}}}}{\Delta }J_x^2,\label{eq14}
	\end{eqnarray}
	where the second term denotes the ac Stark shift of the energy of the two-level system induced by the normal mode $B_-$. We note that, besides the spin nonlinear term $J_x^2$, another nonlinear term $J_z^2$ now also appears. The contribution of these two terms can be freely tuned by varying the modulation magnitude $\xi$. Next, we assume ${J_{{m_0}}}(\xi ) =-{J_0}(\xi ) $, which results in  $ ( {\lambda _0 - \lambda _{{m_0}}} )^2 = -{4\lambda _0\lambda _{{m_0}}}=4\lambda^2 J^2_0(\xi)$, and finally arrive at
	\begin{eqnarray}
		{H_{{\rm{eff}}}} &=& \Delta B_ - ^\dag {B_ - } + \frac{4{\lambda ^2}J_0^2(\xi )}{\Delta}\left( {J_z^2 + J_x^2} \right)\nonumber\\
		&=& \Delta B_ - ^\dag {B_ - } - \epsilon{\chi}J_y^2,\label{eq15}
	\end{eqnarray}
	where $\epsilon=J_0^2(\xi){\omega_-/\Delta}$. The Hamiltonian (\ref{eq15}) is a $y$-axis OAT, whose coupling strength is reduced by a factor $\epsilon$.
	\begin{figure}[t]
		\centering
		\includegraphics[scale=0.48]{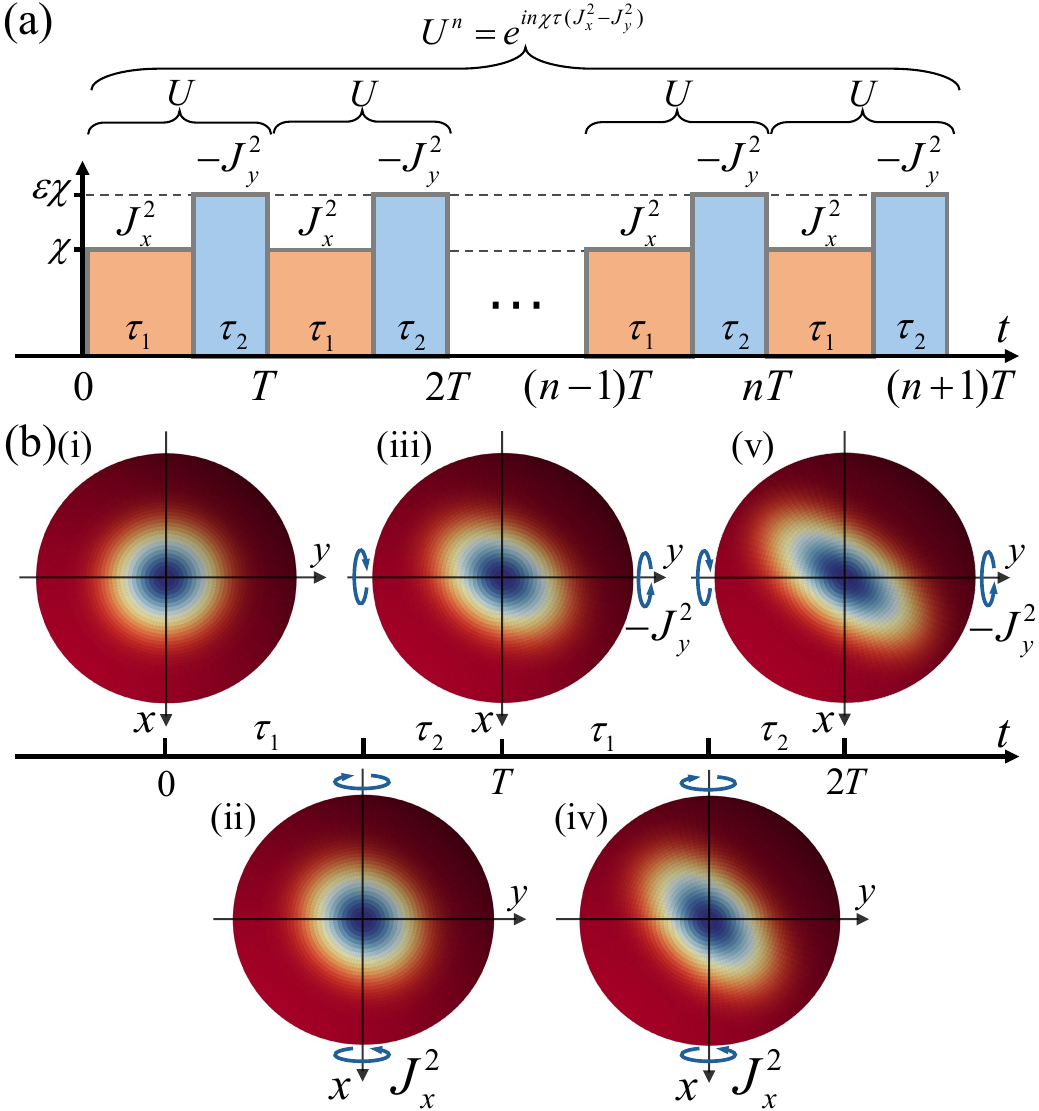}
		\caption{Schematic depiction of the interaction sequence of the proposed TAT scheme vs time $t$ in arbitrary units, and the associated spin distribution on the Bloch sphere for the first two evolutionary periods. (a) The interaction sequence is cyclic with period $T=\tau_1+\tau_2$. The area of a square or a rectangle is proportional to the interaction strength: the squares denote applying $U_x=\exp(i\chi \tau J_x^2)$ to the spins, while the rectangles represent applying $U_y=\exp(-i\chi \tau J_y^2)$ to the spins. (b) (i) shows the initial CSS $\ket{J,J}$, (ii) shows SSS generated by $U_x$, and (iii) shows the SSS generated by the net evolution $U_yU_x$, which is analogy to the SSSs created by the TAT evolution, $U=\exp[i\chi \tau (J_x^2-J_y^2)]$, whose uncertainty is squeezed along the $3\pi/4$ direction in the $xy$ plane. (iv) and (v) exhibit a more clearer trend.}
		\label{fig3}
	\end{figure}
	
We are now equipped to implement the TAT interaction. The spins are first subjected to the time evolution $U_x=\exp(i\chi \tau_1J_x^2)$ according to the interaction of Eq. (\ref{eq7}). After this interaction, the spin modulation is switched on, and then the time evolution is governed by the interaction of Eq. (\ref{eq15}), yielding $U_y=\exp(-i\chi \epsilon\tau_2J_y^2)$, where we assume that $ \epsilon<0 $, which can be achieved through adjusting $ \Delta_B $. If the evolution time $\tau_1,\tau_2$ are short, we obtain
	\begin{eqnarray}
		U = {e^{i\chi {\tau _1} J_x^2}}{e^{ - i\chi \epsilon {\tau _2}J_y^2}} \simeq {e^{  i\chi \tau \left( {J_x^2 - J_y^2} \right)}},
	\end{eqnarray}
	where we have set $\epsilon\tau_2=\tau_1\equiv\tau$. By repeating this procedure $n$ times, the effective evolution would be
	\begin{eqnarray}
		{U^n} \simeq {e^{ in\chi \tau \left( {J_x^2 - J_y^2} \right)}}.\label{eq17}
	\end{eqnarray}
	Obviously, it is exactly the TAT evolution, which enables the production of spin squeezing along the $3\pi/4$ direction in the $yz$ plane \cite{PhysRevA.47.5138}, as shown in \cref{fig3}. It should be mentioned that the OAT interaction can also be transformed into the TAT by making use of multipe $\pi/2$ pulse as shown in Ref. \cite{PhysRevLett.107.013601}. This scheme involves multiple accurate rotation of the macroscopic spin, which, however, is quite a challenge in realistic implementations \cite{PhysRevLett.110.163602}. While our proposed scheme relies only on the engineering of the uncertainty distribution in the $xy$ plane, keeping the macroscopic spin intactly (since the modulation Hamiltonian commutates with the macroscopic spin $J_z$), it simplifies the experimental implementation and thus provides the potential to generate spin squeezing with high degrees.
	
	\section{\label{sec:level3} SPIN SQUEEZING}
	To quantify the degree of spin squeezing generated in this system, we utilize the squeezing parameter introduced by Wineland \cite{PhysRevA.46.R6797,PhysRevA.50.67}:
	\begin{eqnarray}
		{\xi ^2} = \frac{{2J\left\langle {\Delta J_{\min }^2} \right\rangle }}{{{{\left\langle {{J_z}} \right\rangle }^2}}},\label{eq16}
	\end{eqnarray}
	where $\left\langle {\Delta J_{\min }^2} \right\rangle $ is the minimum variance in a direction vertical to the mean spin direction $ J_z $ and $J$ is the total angular momentum. If $ \xi^2 < 1 $, then the state is spin squeezed. The smaller the value of $ \xi ^2 $, the stronger the spin squeezing will be achieved.
	\begin{figure*}[t]
		\centering
		\includegraphics[scale=0.53]{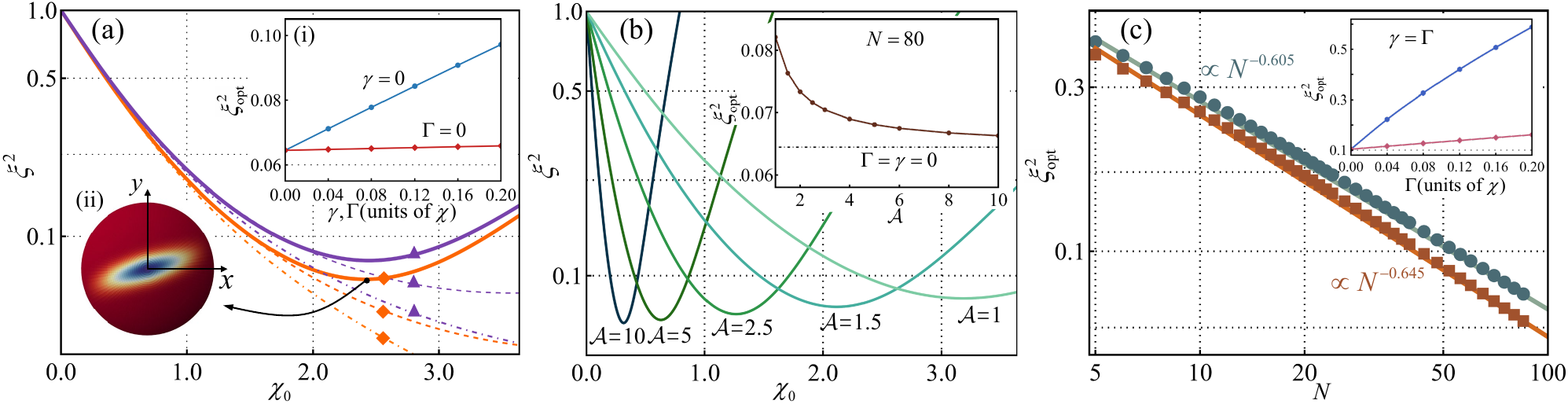}
		\caption{(a) The squeezing parameter $ \xi ^2 $  varies with coupling constant $\chi_0$ under different dissipation conditions. Purple lines with triangle: no loss; orange lines with diamond: $\gamma=0.04$, $\Gamma=0.016$, and $ n_{\text{th}}=2 $. The solid lines represent the exact numerical solution, while the dashed lines correspond to the results from evolution of the expectation values of the angular momentum operators in a second-order cumulant expansion as given Eqs. (\ref{eq:whole1}). The dotdashed lines are the results obtained from the Holstein-Primakoff approximation
			given in Eqs. (\ref{eq23}). The inset (i) shows the maximal achievable spin squeezing vs the single spin dephasing (red line with diamonds) or the collective spin relaxation (blue line with circles); the inset (ii) depicts the optimally one-axis twisted state in the absence of noises on the Bloch sphere. Here we take $N=80$. (b) Performance of the OAT spin squeezing versus $\chi_0$ for different $\mathcal{A}$ under the same decay conditions as (a). Inset shows the optimal squeezing as function of $\mathcal{A}$. The dot-dashed line denotes maximal squeezing achieved by OAT in the absence of noises.  (c) Optimal squeezing parameters $ \xi _{{\rm{opt}}}^2 $ in the presence of different dissipation versus spin number $N$:  $\gamma=0.04,\Gamma=0.08$ (green line with circles) and $\gamma=\Gamma=0.005$ (brown lines with squares). Inset compares the performance of our scheme (red line with diamonds) with the scheme of Ref. \cite{PhysRevLett.110.156402} (blue line with circles) for $N=40$, where we have set $\Gamma=\gamma$.
		}
		\label{fig4}
	\end{figure*}

	We assume that the NV ensemble is initialized along the $z$-axis of the collective Bloch sphere to form the coherent spin state (CSS) $ \left| {{J,J}} \right\rangle $, which is the eigenstate of $J_z$, satisfying $ {J_z}\left| {{J,J}} \right\rangle  = J\left| {{J,J}} \right\rangle $. Such a state has equal transverse variances, $\text{Var}(J_x)=\text{Var}(J_y)=J/2$, where the variance of the operators is defined as $\text{Var}(O)=\langle O^2\rangle-\langle O\rangle^2$ \cite{PhysRevA.47.5138}. During the interaction, the spin system inevitably experiences decoherence. First, the spins undergos single-spin dephasing at a rate $\gamma=1/2T_2$, while the single-spin relaxation as $T_1$ can be neglected due to long decoherence time at low temperatures \cite{PhysRevLett.108.197601}. Second, the dissipation of mechanical and cavity modes also cause the collective decay of the spin at a rate $\Gamma  = {\omega _a}/Q{r^2}$ \cite{PhysRevLett.110.156402}, where $Q$ denotes the quality factor of mechanical oscillator \cite{PhysRevLett.113.020503}. We note that a direct way of circumventing the influence of collective decay is to increase the values of $r$, which, on the other hand, also reduces the coupling constant $\chi$ according to Eq. (\ref{eq8}). Therefore, for a given $N$ there exists an optimal $r$ that can maximize the amount of squeezing.
	After taking all these effects into account, the dynamics of the system under the OAT interaction of Eq. (\ref{eq7}) can be described by the master equation
	\begin{eqnarray}
		\nonumber \frac{\partial\rho}{\partial t} &=& - i\left[ {-\chi J_x^2,\rho } \right] + \gamma \sum\limits_{i} {\mathcal{D}\left[ {\sigma _z^i} \right]\rho }\\
		&& + \Gamma \left( {2{n_{\rm{th}}} + 1} \right)\mathcal{D}\left[ {{J_x}} \right]\rho,\label{eq19}
	\end{eqnarray}
	where $ \mathcal{D}\left[ O \right]\rho  = O\rho {O^\dag } - \frac{1}{2}\left( {{O^\dag }O\rho  + \rho {O^\dag }O} \right) $ and $\bar n_{\rm{th}}$ denotes the mean excitation number of the thermal reservoir. Using Eq. (\ref{eq19}) and the time evolution of the expectation values of
	the spin operators $\partial_t\langle A\rangle=\operatorname{tr}\{A \dot{\rho}\}$, one can derive the time evolution of spin averages and variances
	\begin{subequations}
		\label{eq:whole1}
		\begin{eqnarray}
			{\partial _t}\left\langle {{J_z}} \right\rangle &=& -2\chi \left\langle {{C_{xy}}} \right\rangle  - \frac{1}{2}\Gamma \left( {2{n_{\rm{th}}} + 1} \right)\left\langle {{J_z}} \right\rangle, \label{20a}\\
			{\partial _t}\left\langle {J_x^2} \right\rangle &=& - 4\gamma \left\langle {J_x^2} \right\rangle  + 2\gamma J,\\
			{\partial _t}\left\langle {J_y^2} \right\rangle &=&  4\chi \left\langle {{C_{xy}}} \right\rangle \left\langle {{J_z}} \right\rangle - 4\gamma \left\langle {J_y^2} \right\rangle  + 2\gamma J  \nonumber \\
			&&- \Gamma \left( {2{n_{\text{th}}} + 1} \right)\left( {\left\langle {J_y^2} \right\rangle  - \left\langle {J_z^2} \right\rangle } \right),\\
			{\partial _t}\left\langle {J_z^2} \right\rangle &=& -4\chi \left\langle {{C_{xy}}} \right\rangle \left\langle {{J_z}} \right\rangle \nonumber \\
			&& -\Gamma \left( {2{n_{\text{th}}} + 1} \right)\left( {\left\langle {J_z^2} \right\rangle  - \left\langle {J_y^2} \right\rangle } \right),\\
			{\partial _t}\left\langle {{C_{xy}}} \right\rangle &=&  2\chi \left\langle {J_x^2} \right\rangle \left\langle {{J_z}} \right\rangle  - 4\gamma \left\langle {{C_{xy}}} \right\rangle \nonumber \\
			&& - \frac{1}{2}\Gamma \left( {2{n_{\text{th}}} + 1} \right)\left\langle {{C_{xy}}} \right\rangle.\label{eq20}
		\end{eqnarray}
	\end{subequations}
	where $ {{C_{xy}}} = \left( {{J_x}{J_y} + {J_y}{J_x}} \right)/2 $.
	In deriving Eqs. (\ref{eq:whole1}) we have neglected third-order cumulants by making the approximation $\left\langle {{J_i}{J_j}{J_k}} \right\rangle  \approx {\rm{ }}\left\langle {{J_i}{J_j}} \right\rangle \left\langle {{J_k}} \right\rangle  + \left\langle {{J_i}{J_k}} \right\rangle \left\langle {{J_j}} \right\rangle  + \left\langle {{J_j}{J_k}} \right\rangle \left\langle {{J_i}} \right\rangle  - 2\left\langle {{J_i}} \right\rangle \left\langle {{J_j}} \right\rangle \left\langle {{J_k}} \right\rangle $ \cite{doi:10.1143/JPSJ.17.1100,PhysRevA.100.013856}. Next, we consider the case that the spin number $N$ is large and the macroscopic spin after the OAT interaction remains large, which enables us to apply the approximation $\langle J_z\rangle\approx J$ (also known as the Holstein-Primakoff approximation \cite{PhysRev.58.1098}) to the spin ensemble. Then, Eqs. (\ref{eq:whole1}) can be simplified as
	\begin{subequations}
		\label{eq:whole2}
		\begin{eqnarray}
			{\partial _t}\left\langle {J_x^2} \right\rangle &=& - 4\gamma \left\langle {J_x^2} \right\rangle  + 2\gamma J,\\
			{\partial _t}\left\langle {J_y^2} \right\rangle &=& 4\chi J\left\langle {{C_{xy}}} \right\rangle - \left({2\alpha + 4\gamma} \right)\left\langle {J_y^2} \right\rangle + {c_0},\\
			{\partial _t}\left\langle {{C_{xy}}} \right\rangle &=& 2J\chi \left\langle {J_x^2} \right\rangle - \left( {\alpha  + 4\gamma } \right)\left\langle {{C_{xy}}} \right\rangle,\label{21}
		\end{eqnarray}
	\end{subequations}
	where $\alpha = \Gamma({n_{\text{th}}} + 1/2) $ and ${c_0} = 2{J^2}\Gamma {n_{\text{th}}} + {J^2}\Gamma  + 2\gamma J$. The solutions to these differential equations are
	\begin{subequations}
		\label{eq:whole3}
		\begin{eqnarray}
			\left\langle {J_x^2} \right\rangle &=& \frac{J}{2},\\
			\left\langle {J_y^2} \right\rangle &=& \left( {{c_0} - \frac{{4{\chi ^2}{J^3}}}{{\alpha  + 4\gamma }} } \right)\frac{{1 - {e^{ - \left( {2\alpha  + 4\gamma } \right)t}}}}{{2\alpha  + 4\gamma }} + \frac{J}{2}{e^{ - \left( {2\alpha  + 4\gamma } \right)t}} \nonumber \\
			&&+ \frac{{4{\chi ^2}{J^3}}}{{\alpha  + 4\gamma }}\frac{{{e^{ - \left( {\alpha  + 4\gamma } \right)t}} - {e^{ - \left( {2\alpha  + 4\gamma } \right)t}}}}{\alpha },\\
			\left\langle {{C_{xy}}} \right\rangle &=&  \frac{{\chi {J^2}}}{{\alpha  + 4\gamma }}\left[ {1 - {e^{ - \left( {\alpha  + 4\gamma } \right)t}}} \right].\label{22c}
		\end{eqnarray}
	\end{subequations}
	It is well known that the squeezing (minimal-variance) direction of the OAT scheme varies with time $t$ \cite{PhysRevA.46.R6797}, and the minimal variance in the $xy$ plane can be directly calculated via
	\begin{figure*}[t]
		\centering
		\includegraphics[scale=0.52]{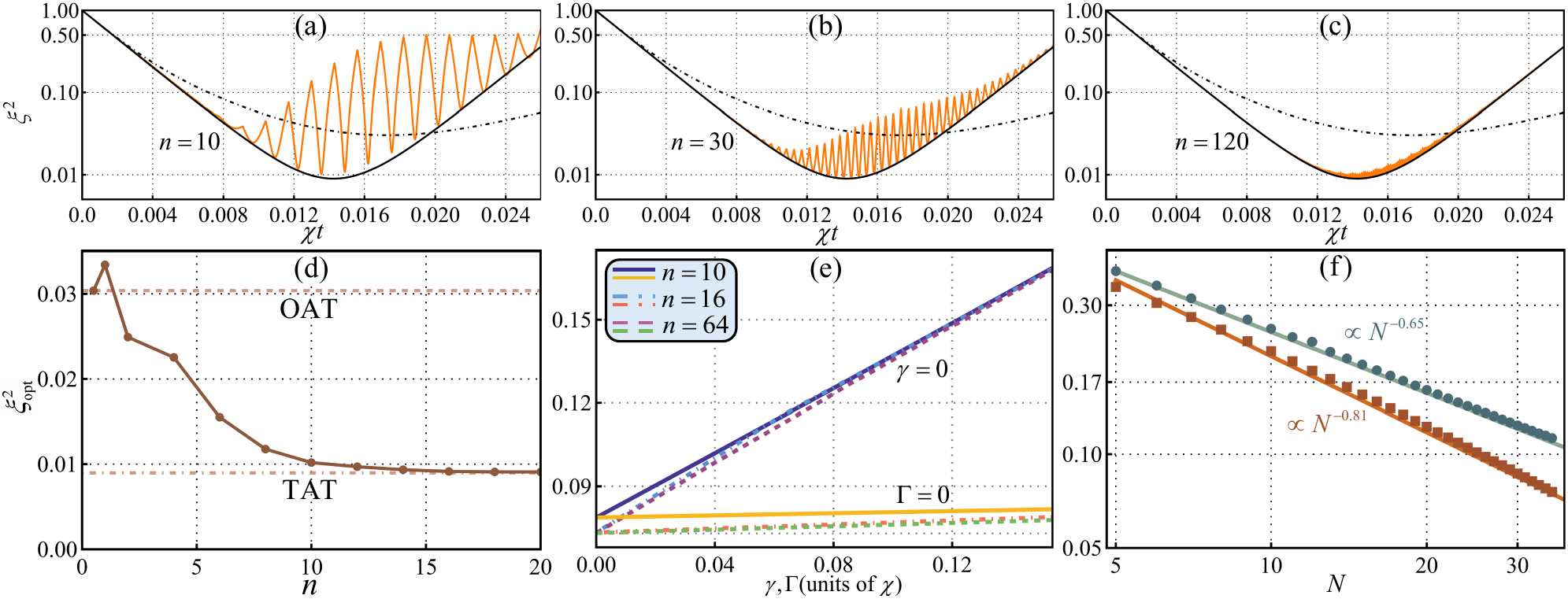}
		\caption{Spin squeezing parameters $\xi^2$ of our proposed TAT scheme (orange solid lines) versus interaction time at different periodic numbers $n=10$ (a), $n=30$ (b), and $n=120$ (c), which are compared with the effective TAT dynamics of Eq. (\ref{eq17}) in black solid
			lines. The black dot-dashed lines denote the OAT dynamic. Here we have taken $N=200$. (d) The maximal achievable spin squeezing of the proposed TAT scheme versus periodic numbers $n$. (e) Performance of the TAT scheme in the presence of either single-spin dephasing (horizontal  lines) or collective-spin relaxation (diagonal lines) for different $n$. (f) Optimal squeezing parameters $ \xi _{{\rm{opt}}}^2 $ in the presence of dissipation versus spin number $N$: $\gamma=0.04$, $\Gamma=0.08$ (green line with circles) and $\gamma=\Gamma=0.005$ (brown line with squares).
		}
		\label{fig5}
	\end{figure*}
	$\left\langle {\Delta J_{\min }^2} \right\rangle = (\langle {J_x^2 + J_y^2} \rangle  - \sqrt {{{ {\langle {J_x^2 - J_y^2} \rangle }}^2} + 4{{\langle {{C_{xy}}} \rangle }^2}})/2$  \cite{MA201189}. Finally, using the results of Eqs. (\ref{eq:whole3}) we get the amount of squeezing according to Eq. (\ref{eq16}) as
	\begin{eqnarray}
		{\xi ^2} = \frac{1}{J}\left\{ {{{\cal V}_ + } - \sqrt {{\cal V}_ - ^2 + 4{\chi ^2}{J^4}{{\left[ {\frac{{1 - {e^{ - \left( {\alpha  + 4\gamma } \right)t}}}}{{\alpha  + 4\gamma }}} \right]}^2}} } \right\},\label{eq23}
	\end{eqnarray}
	where ${\mathcal{V}_ \pm }= \{ J \pm [J\exp [ - (2\kappa  + 4\gamma )t] + 2{c_0}t - 2(2{\chi ^2}{J^3} + \alpha {c_0} + 2\gamma {c_0}){t^2}]\} /2$.
	The result (\ref{eq23}) can be further simplified in the limit of strong coupling strength  $\chi_0=J \chi t \gg 1$ and small spin decays $
	J\Gamma {n_{\text{th}}}t \ll 1,\gamma t \ll 1$. With these approximations the spin squeezing of Eq. (\ref{eq23}) can be expressed as
	\begin{eqnarray}
		{\xi ^2}= \frac{1}{{4{\chi ^2}{J^2}{t^2}}} + \frac{t}{2}\left[ {\Gamma \left( {2{n_{\rm{th}}} + 1} \right) + 4\gamma } \right],\label{eq24}
	\end{eqnarray}
	where the first term denotes the contribution of the unitary OAT squeezing process, while the remaining terms represent the noises added by the two decoherence sources. The competition between the two process leads to an optimal spin-squeezing. Finally, optimizing $ \xi ^2 $ with respect to $ t $ we obtain the optimal squeezing parameter
	\begin{eqnarray}
		\xi _{{\rm{opt}}}^2 = \left( {\frac{1}{{{2^{1/3}}}} + \frac{1}{{{4^{2/3}}}}} \right){\left( {\frac{{\Gamma {n_{\text{th}}} + \Gamma/2  + 2\gamma }}{{\chi J}}} \right)^{2/3}},\label{eq25}
	\end{eqnarray}
	at $t_{\rm{opt}} = 1/{[ {2{\chi ^2}{J^2}( {\Gamma {n_{\text{th}}} + \Gamma/2  + 2\gamma } )} ]^{1/3}}$. The obtained scaling $\xi _{{\rm{opt}}}^2\propto J^{-2/3}$ is similar to the result for spin systems given in Ref. \cite{PhysRevA.47.5138}. In \cref{fig4}(a), we compare the spin squeezing dynamics obtained through a second-order cumulant expansion of Eqs. (\ref{eq:whole1}) and the Holstein-Primakoff approximation of Eqs. (\ref{eq:whole3}) with the exact numerical solution for $N=80$ in the presence and absence of noise effects, which verified that our approximation results agree well with exact results for small coupling constant $\chi_0$. The performance of the scheme in the presence of either collective-spin or single-spin dissipations is also shown in the inset of \cref{fig4}(a), indicating that the scheme is robust against the single-spin dephasing while is sensitivity to the collective decay.
	
	As we mentioned previously that our proposed schemes also offer the potential to enhance the spin-spin coupling by increasing the amplification factor $\mathcal{A}$. Such enhancement relies on the increase of the ratio $\omega_a/g$ and thus is independent on $r$. This means that one can enhance spin-spin coherent interactions while, at the same time, keep the (unwanted) incoherent process (that is, $\Gamma$) unchanged. From equation (\ref{eq25}) we see immediately that the increase of $\chi$ will benefit the achievable spin squeezing. In \cref{fig4}(b), we plot spin squeezing dynamics for different coupling strength $\chi$, showing that the enhancement of coupling strength not only improves but also speeds up the generation of spin squeezing greatly since the optimal-squeezing preparation time $t_{\rm{opt}}\propto\chi^{-2/3}$, which would benefit the realistic applications \cite{barberena2022fast}. As shown in the inset of \cref{fig4}(b), the larger the coupling strength, the smaller the influences of the noise effects.
	
	Figure \ref{fig4}(c) shows optimal squeezing versus number of spins in the presence of different decay. One can observe that, firstly, the squeezing scalings are quite favorable, and secondly, our proposed scheme is highly robust against noise. Even with a significant change in the decoherence rates, the amount of spin squeezing only decreases slightly. The later property can also be further verified by comparing the performance of our scheme with the scheme presented in Ref. \cite{PhysRevLett.110.156402}, as shown in the insets of \cref{fig4}(c). With increasing of either the collective decay $\Gamma$ or the single-spin dephasing $\gamma$, the spin squeezing of the scheme in Ref. \cite{PhysRevLett.110.156402} decreases rapidly, while the noises only have a little impact on our scheme. Such results can be understood intuitively as follows: The OAT evolution leads to the redistribution of uncertainties in the $x$-$y$ plane, as shown in the inset (ii) of \cref{fig4}(a), from which we can see that most of the quasiprobability distribution are along (or near) the equator of the Bloch sphere. Since the CSSs along the equator are very close to the \emph{dark state} $\ket{0_x}$ of $J_x$ [where $\ket{0_x}=\exp(-i\pi J_y/2)\ket{0}$, satisfying $J_x\ket{0_x}=0$], they (and thus the produced squeezed states) are insensitive to collective decoherences. The single spin dephasing preserves the macroscopic spin (since $[\sigma_z,J_z]=0$), which would also benefit the amount of spin squeezing according to Eq. (\ref{eq16}). It is these characteristics that make the present scheme robust against decoherence.
	
	For the TAT scheme described above, the alternate applying of the interactions (\ref{eq7}) and (\ref{eq15}) to the NV centres yields the following master equation
	\begin{eqnarray}
		\frac{{\partial \rho }}{{\partial t}} &=& \mathcal{B}\left( t,0 \right)\left\{ { - i\left[ -{\chi J_x^2,\rho } \right] + \Gamma \left( {2{n_{{\rm{th}}}} + 1} \right){\cal D}\left[ {{J_x}} \right]\rho } \right\}\nonumber\\
		&&\mathcal{B}\left( {t - \tau,\tau } \right)\left\{ { - i\left[ {\epsilon\chi J_y^2,\rho } \right] + \Gamma \left( {2{n_{{\rm{th}}}} + 1} \right){\cal D}\left[ {{J_y}} \right]\rho } \right\}\nonumber\\
		&&+ \gamma \sum\limits_i {{\cal D}\left[ {\sigma _z^i} \right]\rho } ,\label{eq26}
	\end{eqnarray}
	where the periodical box function $\mathcal{B}(t,t_1)=\sum_{m=0}^{n}\{\Theta[t-mT]-\Theta[t-\tau-(1/\epsilon-1)t_1-mT]\}$ with $\Theta$ being  the Heaviside step function. To check the validity of our proposed TAT scheme, we compare the effective evolution of Eq. (\ref{eq17}) with the actual dynamics induced by Eq. (\ref{eq26}) in the absence of decoherence. For fixed $\chi$ and value of $n\chi\tau$ ($\equiv\kappa_0$), we have $n=\kappa_0/\chi\tau$, indicating that one may adjust the number of periodicity $n$ by changing the value of $\tau$. \cref{fig5}(a-c) shows that, as $n$ increases, the evolution induced by our scheme gets more and more close to the ideal TAT evolution. In \cref{fig5}(d), we also study the maximal achievable spin squeezing of the proposed scheme as a function of $n$, showing that our approach outperforms the OAT even at small value of $n$ ($n\geq 3$). A value of $n$
	larger than $10$ allows us to approach the level of spin squeezing achieved by the TAT method. \cref{fig5}(e) shows how our TAT schemes
	performs for different degrees of imperfection, revealing that, analogous to the OAT, the TAT scheme is also much more sensitivity to collective decay. Moreover, further increase of $n$ will not benefit the spin squeezing in the case of large decoherence. To quantify the effectiveness of the TAT protocol, we numerically calculate
	the maximal squeezing under the same imperfection presented in the OAT above as a function of atom number $N$, as plotted in \cref{fig5}(f). The fit of the proposed TAT protocol indicates that a maximum-squeezing scaling $\propto N^{0.81}$ is achievable, which is much better than the ideal OAT scaling $\propto N^{0.67}$ \cite{PhysRevA.47.5138}.
		
\vbox{}
\section{\label{sec:level4} Conclusions}
	In summary, we have proposed a method to generate spin squeezed states of an NV ensemble in a hybrid quantum system. The spin squeezing is achieved by coupling the NV centers via strain to the vibrational mode of a mechanical resonator driven by an optical cavity. In contrast to the previous phonon-induced spin squeezing \cite{PhysRevLett.114.093602}, our phonon-photon induced spin squeezing exhibit many advantages. First, \emph{unitary} OAT evolution of the NV ensemble can be realized, that is, one can completely disentangle the atomic variables from the Boson modes which media the interatomic interaction. As a result, no decoupling operation is required \cite{PhysRevLett.105.093602}, which simplifies the experimental implementation and thus improves the spin squeezing performance \cite{PhysRevA.85.013803}. Second, under some circumstances our proposed scheme also has the potential to realize more stronger spin-spin nonlinear interaction. Third, the spin squeezing approach is robust against decoherence. These good characteristics not only make our scheme suitable for spin squeezing, but also provide a potential for creating atomic GHZ states through the OAT evolution \cite{PhysRevLett.82.1835}. By introducing a periodical modulation to the transition frequency of the NV system, we are able to transfer the OAT scheme into the more efficient TAT scheme, capable of producing spin squeezing near the Heisenberg limit even in the presence of imperfection. Our proposed scheme can also be extended to other quantum platforms such as neutral atoms in a cavity-optomechanics system \cite{frisk2019ultrastrong,aspelmeyer2014cavity}.
	
	\begin{acknowledgments}
		This work was supported by the Natural Science Foundation of China (Grants No. 22273067), the Natural Science Foundation of Zhejiang province, China (Grant No. LQ23A040001, No. LY24A040004), and the Department of Education of Zhejiang Province, China (Grant No. Y202146469).
		
	\end{acknowledgments}
	
	\appendix

	\bibliography{ref}

\begin{thebibliography}{89}
\bibitem [{foo()}]{footnote1}%
  \BibitemOpen
  \href@noop {} {\bibinfo  {journal} {The coupling strength given in Ref.
  \cite{PhysRevLett.110.156402} is $g^2/\Delta$, where $\Delta$ is the detuning
  between the mechanical mode and NV centers. Similarly, the coupling strength
  is also obtained under the condition $\Delta \gg g$. So for comparision, we
  also assumed that $r=\Delta/g\gg 1$}\ }\BibitemShut {NoStop}%
\end{thebibliography}
	
\end{document}